\documentclass[a4paper]{article}
\usepackage[english]{babel}
\usepackage{amssymb, amsmath}
\usepackage{amsfonts}
\usepackage{hyperref}
\usepackage[utf8]{inputenc}
\usepackage[T1]{fontenc}
\usepackage[usenames,dvipsnames]{color}
\usepackage[textsize=footnotesize]{todonotes}
\usepackage[vlined, ruled, linesnumbered]{algorithm2e}
\usetikzlibrary{intersections,positioning,arrows}

\DeclareMathOperator{\dist}{dist}
\DontPrintSemicolon

\begin{document}
\title{Massively Parallel Construction of the Cell Graph
\footnote{The research funded by National Science Center, decision  DEC-2012/07/D/ST6/02483.}
}

\author{
Krzysztof Kaczmarski, Pawe{\l} Rz{\k a}{\. z}ewski  \\
Albert Wolant\\
\texttt{\{k.kaczmarski,p.rzazewski\}@mini.pw.edu.pl} \\
\texttt{wolanta@student.mini.pw.edu.pl} \\ \\
Warsaw University of Technology \\
Faculty of Mathematics and Information Science,\\
Koszykowa 75 , 00-662 Warsaw, Poland
}
\date{ }
\maketitle

\begin{abstract}
Motion planning is an important and well-studied field of robotics. A typical approach to finding a route is to construct a {\em cell graph} representing a scene and then to find a path in such a graph. 
In this paper we present and analyze parallel algorithms for constructing the cell graph on a SIMD-like GPU processor.

Additionally, we present a new implementation of the dictionary data type on a GPU device. In the contrary to hash tables, which are common in GPU algorithms, it uses a search tree in which all values are kept in leaves. With such a structure we can effectively perform dictionary operations on a set of long vectors over a limited alphabet.
\end{abstract}

\maketitle

\section{Introduction}

Motion planning is a common task in robotics and artificial intelligence. One of the aims is to find a path, which can be traversed by a rigid body (e.g. a robot) to get to the destination point and avoid collisions with obstacles \cite{Hwang}.
Dobrowolski \cite{DBLP:conf/sii/Dobrowolski13} considered the problem of motion planning in $SO(3)$ space (i.e. rotations about the origin in the Euclidean 3-space). He presented algorithms for constructing the {\em cell graph}, i.e. a graph representation of the configuration space. It is worth noting that these algorithms work for any space, not only for $SO(3)$. 
Although the algorithms presented by Dobrowolski proved to be significantly faster than the naive approach, their running time was not acceptable for complicated scenes. As the main contribution of this paper, we present and analyze parallel extensions of these algorithms for GPU processors. 
%

One of the parallel algorithms introduced in Section~\ref{sec:parallel} uses some variation of a \textit{binary search tree}, in which all elements are kept in leaves. There are several known implementations of a binary search tree on the GPGPU (see for example \cite{Kim:2010:FFA:1807167.1807206}). 
Our implementation allows us for an efficient execution of dictionary operations on a set of long vectors over an alphabet of a constant size. 
As a dictionary is a fundamental data type, widely used in many applications, we believe that our solution may be interesting and important on its own.


\subsection{Definitions and basic properties}
Let $n, \ell \in \mathbb{N}$. By $[n]$ we denote the set $\{0,1,\ldots,n-1\}$. 
By $[n]^\ell$ we denote the set of all vectors of length $\ell$ over the alphabet $[n]$. The $i$-th coordinate (for binary vectors called the {\em $i$-th bit}) of a vector $x$ is denoted by $x(i)$. The coordinates are indexed in zero-based convention, i.e. $x = x(0),x(1),\ldots,x(\ell-1)$. For $i,j$ such that $0 \leq i < j < \ell$, by $x(i;j)$ we denote the segment $x(i),x(i+1),\ldots,x(j-1)$. 

The {\em Hamming distance} of two binary vectors $x,y \in [2]^\ell$, denoted by $\dist(x,y)$, is the number of positions $i$, such that $x(i) \neq y(i)$. Observe that $\dist$ is a metric function, so it satisfies the triangle inequality: $\dist(x,y) + \dist(y,z) \geq \dist(x,z)$.
From this it follows that: $(\star)~~\dist(x,y) \geq |\dist(x,z) - \dist(y,z)|$.

\section{Problem of Cell Graph Construction for Motion Planning} 
\label{sec:motivation}

In this section we describe the notion of the cell graph in motion planning. Although we use a very simple example, similar methods can be (and actually are) used in much more complicated settings (see for example \cite{Canny, Hwang}).

\subsection{Cells and vectors}

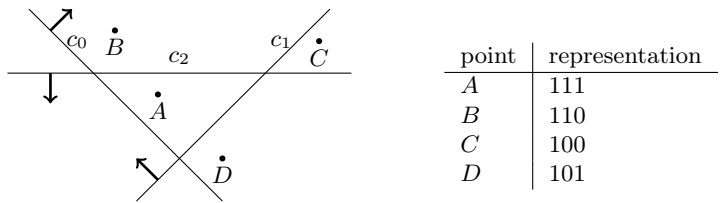
\begin{figure}[htb]
\small
\begin{center}
\raisebox{-30pt}
{
\begin{tikzpicture}[scale = 0.4, rotate = 45]
\draw (5,0) -- node[above, near end] {$c_0$} (5,9);
\draw (3,2) -- node[near end, above] {$c_1$} (12,2);
\draw (3,8) -- node[left, above] {$c_2$} (11,0);
\draw[line width = 1,->] (5,8) -- (6,8);
\draw[line width = 1,->] (4,2) -- (4,3);
\draw[line width = 1,->] (4,7) -- (4-0.71,7-0.71);
\fill (6,4) circle (0.1) node[below]{$A$};
\fill (6.5,6.5) circle (0.1) node[below]{$B$};
\fill (11,1.5) circle (0.1) node[below]{$C$};
\fill (6,1) circle (0.1) node[below]{$D$};
\end{tikzpicture}
}
\hskip 1cm
\begin{tabular}{l | l}
point & representation \\ \hline
$A$ & 111 \\
$B$ & 110 \\
$C$ & 100 \\
$D$ & 101
\end{tabular}
\end{center}
\caption{The arrangement of lines partitions the space into cells. Our object is given by three linear inequalities.
}
\label{fig-lines}
\end{figure}

Let us consider a system of inequalities $c_0,c_1,\ldots,c_{\ell-1}$ (constraints), describing the boundaries (see Figure \ref{fig-lines}), that
partition the space into a number of pairwise disjoint regions, called {\em cells}. We say that two cells are neighboring (a robot can move directly from one to another) if their boundaries share some arc (one point is not enough). Our task is to unify the cells and say which of them are neighbors.
Then an obstacle-free route for a robot can be determined using a graph algorithm (see for example \cite{1664021}). As the scenes (i.e. the space with the arrangement of obstacles) in real-life applications tend to be very complicated, an effective construction of the cell graph is a crucial part of this approach.
 

We shall represent each point $P \in \mathbb{R}^2$ by an $\ell$-element binary vector $v_P$. The $i$-th bit of $v_P$ is 1 iff $P$ satisfies the inequality $c_i$ (see Figure \ref{fig-lines}). 
Observe that such a representation identifies all points belonging to the same cell, so it can be seen as a representation of this cell. The neighboring cells are exactly the cells whose representants differ on exactly one position, which means that their Hamming distance is 1.
There are several known approaches to parallel computation of the Hamming distance in various settings \cite{DBLP:conf/ipps/LiuGLRL12, DBLP:conf/iros/PanLM10, Grabowski2008182}. However, none of them benefits for the specific properties of our task.



Observe that the theoretical bound for the number of different cells in the scene with $\ell$ constraints is $2^\ell$. However, due to the arrangement of constraints, the actual number of non-empty cells is usually much smaller. For example, there is no point represented by a vector $000$ in Figure \ref{fig-lines}.
This is the reason why detecting all cells for the given scene is a hard task. However, we are usually satisfied by approximate solutions generated by a randomized procedure called a {\em sample generator}. This procedure generates (and possibly accumulates) some sequence of random points. The simplest one is the so-called {\em Shoemake's method} \cite{Shoemake:1992:URR:130745.130769}, in which the random points are generated uniformly.

\subsection{Constructing the cell graph}
Let $X = \{x_0,x_1,\ldots,x_{n-1}\}$ be a set of $n$ binary vectors, each of length $\ell$. These vectors represent the cells and are generated by a sample generator.
The {\em cell graph} $G_X$ for $X$ is the graph with vertex set $X$, in which edges are all pairs of vectors $x_i, x_j$ (for $0 \leq i < j < n$), such that $\dist(x_i,x_j) = 1$.

In this paper we are interested in solving the problem of constructing the cell graph, i.e. \textbf{finding the edge set of $G_X$} for the \textbf{given set $X \subseteq [2]^\ell$}.
Since the degree of each vertex is at most $\ell$, the graph $G_X$ has at most $\frac{n \cdot \ell}{2}$ edges. Moreover, the total number of bits in $X$ is $n \cdot \ell$.  Thus the lower bound for the complexity of any algorithm constructing the cell graph is $\Omega(n \cdot \ell)$.

It is also worth mentioning that many sample generators used in applications are not perfect and may output some vector more than once (so $X$ is in fact a multiset). Observe that in this case the number of pairs $i,j$ such that $\dist(x_i,x_j)=1$ can  increase to $\Theta(n^2)$.
A desired property of any algorithm constructing the cell graph is to be able to deal with such a situation and output only unique pairs of neighboring vectors, without increasing the complexity.

When comparing the complexities of the algorithms we will assume that $\ell \ll n$. This is justified, since in most practical applications $n$ is about a few million, while $\ell$ is about a few hundreds.

\section{Parallel  algorithms} \label{sec:parallel}

Parallel algorithms presented in this section are inspired by sequential algorithms for the problem presented by Dobrowolski \cite{DBLP:conf/sii/Dobrowolski13}. 

\subsection{Heuristic algorithm}

In the naive approach we compare all pairs of vectors in total time $O(n^2 \cdot \ell)$ \cite{DBLP:conf/sii/Dobrowolski13}. We improve this method by choosing a small constant $h \in \mathbb{N}$ and computing the distance between each of vectors $x_0,x_1,\ldots,x_{h-1}$ and each vector in $X$.
Then, for each pair of vectors we compare them with each other to determine if their Hamming distance is 1. We are able to discard some pairs faster, using formula $(\star)$ and previously computed distances to vectors $x_0,x_1,\ldots,x_{h-1}$. Observe that for $h=0$ this algorithm reduces to the naive one.

Each pair of vectors $x_i,x_j$ is considered by one block of $w$ threads. Each thread from this block considers the pair of corresponding segments of vectors $x_i$ and $x_j$.
For simplicity, assume that $w$ divides $\ell$ (otherwise the last thread of the block works on shorter segments). Each thread in the block compares a $\left(\frac{\ell}{w}\right)$-element segment of $x_i$ with the corresponding segment of $x_j$. More specifically, the $t$-th thread (for $t \in [w]$) of the block operates on segments $x_i(t \cdot \frac{\ell}{w} ; (t+1) \frac{\ell}{w})$.

First we compute $\dist(x_i,x_j)$ for $i \in [h]$ and $j \in [n]$ and store the values in a shared memory.
Each thread writes the number of positions, on which its segments differ, to a single cell of the shared vector $results$.
Then all those values are summed in parallel.  
Algorithm \ref{alg:par-compute-dist} shows the pseudo-code of this step.

\begin{algorithm}[htb]
\small
\caption {ComputeDist}
\label{alg:par-compute-dist}
\SetKw{Print}{output}
\SetKw{Continue}{Continue}
\SetKw{Break}{Break}
\SetKwFor{BPFor}{for}{do in parallel (blocks)}{}
\SetKwFor{TPFor}{for}{do in parallel (threads)}{}

\KwIn{$X = \{x_0,x_1,\ldots,x_{n-1}\} \subseteq [2]^\ell, h \in \mathbb{N}$}
initialize $dist(x_i,x_j) = 0$ for all $i \in [h], j \in [n]$\\
\BPFor {$i \in [h]$ and $j \in \{i+1,\ldots,n-1\}$}
{	
	initialize $results[t] = 0$ for all $t \in [w]$\\
	\TPFor {$t \in [w]$}
	{		
		\For{$k \gets t \cdot \frac{\ell}{w}$ \KwTo $(t+1) \cdot \frac{\ell}{w}-1$}
		{
			\lIf {$x_i(k) \neq x_j(k)$}{$results[t] \gets results[t]+1$}			
		}
		
	}
	$dist(x_i,x_j) \gets \sum_{t\in[w]} results[t]$\\
}
\end{algorithm}

The remaining part, shown in Algorithm \ref{alg:par-heuristic}, is analogous. The difference is that we may stop if we discover that it is greater than 1.

\begin{algorithm}[htb]
\small
\caption {ParallelHeuristic}
\label{alg:par-heuristic}
\SetKw{Print}{output}
\SetKw{Continue}{Continue}
\SetKw{Break}{Break}
\SetKwFor{BPFor}{for}{do in parallel (blocks)}{}
\SetKwFor{TPFor}{for}{do in parallel (threads)}{}

\KwIn{$X = \{x_1,x_2,\ldots,x_n\} \subseteq [2]^\ell, h \in \mathbb{N}$}
$dist \gets ComputeDist(X,h)$\\
$results \gets$ vector of $w$ zeros\\

\BPFor {$ h \leq i \leq n-1$ and $i < j \leq n-1$}
{
	\If{$|\dist(x_i,x_d) - \dist(x_j,x_d)| > 1$ for all $d \in [h]$}
	{
		\TPFor {$t \in [w]$}
		{
		$c \gets 0$\\
		\For{$k \gets t \cdot \frac{\ell}{w}$ \KwTo $(t+1) \cdot \frac{\ell}{w}-1$}
		{
			\lIf {$x_i(k) \neq x_j(k)$}{$c \gets c+1$}
			\lIf {$c \geq 2$}{\Break	}
		}
		$results[t] \gets c$\\
		}
		$count \gets 0$\\
	\For{$t \in [w]$}
	{
		$count \gets count + results(t)$\\
		\lIf {$count \geq 2$}{\Break	}
	}
	\lIf{$count = 1$}{\Print $(x_i,x_j)$}
		}
	}			
\end{algorithm}

The worst-case time complexity of this algorithm is $\Theta(n^2 \cdot h \cdot \ell)$, while the space complexity is $\Theta(h \cdot n)$, as we need to store the values of $\dist(x_i,x_j)$. Since $h$ is chosen to be a constant, the time complexity and the space complexity are $\Theta(n^2 \cdot \ell)$ and $\Theta(n)$, respectively. 
Observe that the choice of $h$ strongly affects the constants in the bounds for the complexity. However, the experiments show that even if $h$ is small, the effect on execution time may be significant.

\subsection{Tree-based algorithm}

The main drawback of the previous approach is that it is not aware of the structure of the constructed cell graph. 
Thus Dobrowolski \cite{DBLP:conf/sii/Dobrowolski13} presented an optimized algorithm, based on a different approach. This algorithm first constructs an auxiliary binary tree, storing all vectors in $X$. Using this tree we can determine if the particular vector $x$ is in $X$ in time $O(\ell)$.

\begin{figure}[ht]
\centering
\begin{center}
\raisebox{-30pt}
{
\tiny
\begin{tikzpicture}[scale=0.7]
\tikzstyle{nnode}=[draw, shape = circle]

\node[nnode] (000) at (0,0) {};
\node[fill = black, nnode] (021) at (1,0) {};
\node[nnode] (123) at (2,0) {};
\node[nnode] (310) at (3,0) {};
\node[nnode] (312) at (4,0) {};
\node[nnode] (313) at (5,0) {};

\node[nnode] (00) at (0,1) {};
\node[fill = black, nnode] (02) at (1,1) {};
\node[nnode] (12) at (2,1) {};
\node[nnode] (31) at (4,1) {};

\node[fill = black, nnode] (0) at (0.5,2) {};
\node[nnode] (1) at (2,2) {};
\node[nnode] (3) at (4,2) {};

\node[fill = black,nnode] (root) at (2.5,3) {};

\draw[line width=1] (root) -- (0) node [left, midway] {0};
\draw (root) -- (1) node [right, midway] {1};
\draw (root) -- (3) node [right, midway] {3};

\draw (0) -- (00) node [left, midway] {0};
\draw[line width=1] (0) -- (02) node [left, midway] {2};
\draw (1) -- (12) node [right, midway] {2};
\draw (3) -- (31) node [right, midway] {1};

\draw (00) -- (000) node [left, midway] {0};
\draw[line width=1] (02) -- (021) node [left, midway] {1};
\draw (12) -- (123) node [left, midway] {3};
\draw (31) -- (310) node [right, midway] {0};
\draw (31) -- (312) node [right, midway] {2};
\draw (31) -- (313) node [right, midway] {3};
\end{tikzpicture}
}
\small
\hskip 1cm
\begin{tabular}{l|l}
$x$ & $\widetilde x$ \\ \hline
000000 & 000 \\
001001 & 021 \\
011011 & 123 \\
110100 & 310 \\
110110 & 312 \\
110111 & 313 
\end{tabular}
\end{center}
\caption{A search tree for $r=2$. The search path for $\widetilde{x} = 021$ is marked.} 
\label{fig:r-tree}
\end{figure}
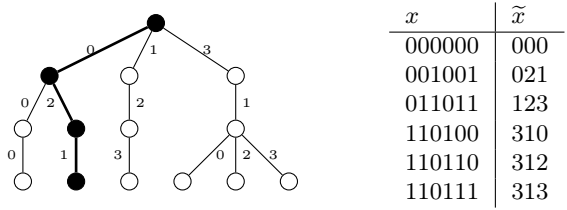

In the parallel version of this algorithm, to improve memory accesses (see Section \ref{sec:implissues}) we use a $2^r$-ary tree $T$ for $r \geq 1$. Let $r$ be fixed and suppose for simplicity that $r$ divides $\ell$ (otherwise the last segment of each vector is considered in a slightly different way).
For $x \in X$, let $\widetilde x$ denote the vector in $[2^r]^{\ell/r}$ such that for every $i \in [\ell/r]$ the sequence
$x(i \cdot \frac{\ell}{r} ; (i+1) \cdot \frac{\ell}{r}-1)$ is the binary encoding of $\widetilde x(i)$. By $\widetilde X$ we denote the set $\{\widetilde x \colon x \in X\}$. We can see $T$ as the representation of the sequences in  $\widetilde X$. 
The tree $T$ has $\ell/r$ levels. Each level of $T$ corresponds to $i$-th coordinate of $\widetilde x$.
Each node contains $2^r$ pointers to nodes of the next level, each corresponding to a different element to $[2^r]$ (see Figure \ref{fig:r-tree} for an example).
If a particular child does not exist, then there are no vectors with the particular prefix. If $C$ is a node of $T$ and $C'$ is its child node, corresponding to the value $v \in [2^r]$, then we say that $C'$ is a {\em $v$-child} of $C$.

The first step of our algorithm is sorting the vectors in $X$. As these vectors are binary, the sorting can clearly be done in $O(n \cdot \ell)$ time, using the {\em radix sort} algorithm.
During this step we also remove all duplicates in $X$.

Then we proceed to constructing the search tree $T$.
Each level of $T$ is constructed in parallel, with synchronization of threads after finishing each level.
For each node $C$ and $v \in [2^r]$ we introduce the set $vectors(C, v)$. Let $C$ be a node on level $h$.
The set $vectors(C,v)$ consists of vectors $\widetilde x \in \widetilde X$, such that: i) $\widetilde x(0 ; h-1)$ is represented by $C$ in $T$ (with just a little abuse of notation we assume that for $h=0$ every vector satisfies this condition), and ii) $\widetilde x (h) = v$.
This means that the vectors from $vectors(C,v)$ are exactly the ones, whose search path begins with the path from the root to the $v$-th child of $C$. Observe that since the set $X$ (and thus $\widetilde X$) is sorted, each set $vectors(C,v)$ can be represented by just two indices -- of the first and of the last vector from this set.
The Algorithm \ref{alg:par-construct} shows the pseudo-code for this step. Observe that the computational complexity of Algorithm \ref{alg:par-construct} is $O(n \cdot \ell)$ (recall that $r$ is a constant).

\begin{algorithm}[htb]
\small
\caption {ConstructTree}
\label{alg:par-construct}
\SetKw{Print}{output}
\SetKw{Continue}{Continue}
\SetKw{Break}{Break}
\SetKwFor{BPFor}{for}{do in parallel (blocks)}{}
\SetKwFor{TPFor}{for}{do in parallel (threads)}{}
\KwIn{$\widetilde X = \{\widetilde x_0,\widetilde x_1,\ldots,\widetilde x_{n-1}\} \subseteq [2^r]^{\ell/r}$}
create the root node (on level 0)\\
\lForEach{$\widetilde x \in \widetilde X$}
		{
			add $\widetilde x$ to the set $vectors(root, \widetilde x(0))$
		}

\For{$h \gets 1$ \KwTo $\ell/r-1$}
{
\TPFor{node $C$ in level $h-1$ and $v \in [2^r]$}
{
	\If {$vectors(C, v) \neq \emptyset$}
	{
		create node $C'$, being the $v$-child of $C$\\
		\lForEach{$\widetilde x \in vectors(C,v)$}
		{
			add $\widetilde x$ to the set $vectors(C', \widetilde x(h))$
		}	
	}	
}
}

\end{algorithm}

After constructing the search tree, we can proceed to the main step -- identifying neighbors.
For every vector in $x \in X$ and every possible neighbor $x'$ of $x$ we check if $x' \in X$ (in fact we check is $\widetilde{x'} \in \widetilde X$). Again, we do it in parallel.
For every vector $\widetilde x$, each bit of $\widetilde x$ is considered by a separate thread.
Observe that each bit of $\widetilde x$ corresponds to a single potential neighbor of $\widetilde x$. Thus each thread checks if this potential neighbor exists. The Algorithm \ref{alg:par-treebased} shows the pseudo-code of this procedure.

\begin{algorithm}[htb]
\small
\caption {ParallelTreeBased}
\label{alg:par-treebased}
\SetKw{Print}{output}
\SetKw{Continue}{Continue}
\SetKw{Break}{Break}
\SetKw{Stop}{Exit thread }
\SetKwFor{BPFor}{for}{do in parallel (blocks)}{}
\SetKwFor{TPFor}{for}{do in parallel (threads)}{}
\KwIn{$X = \{x_0,x_1,\ldots,x_{n-1}\} \subseteq [2]^{\ell}$}
sort $X$\\
$T \gets ConstructTree(\widetilde X)$\\
\BPFor{$x \in X$}
{
	\TPFor{$k \in [\ell]$}
	{
	$x' \gets$ $x$ with the $k$-th bit negated\\
	$C \gets$ the root of $T$\\
	\For {$h \gets 0$ \KwTo $\ell/r-1$}
	{
		$v \gets \widetilde{x'}(h)$\\
		\lIf {there is no $v$-child of $C$} {\Stop}
		$C \gets v$-child of $C$\\
	}
		\Print {$(x, x')$}
	}
}
\end{algorithm}

Observe that we do not have to keep the vector $x'$ explicitly. At each step we need a segment of $x'$ (corresponding to the current level of $T$), which can be found in constant time.
The time complexity of the searching procedure is $O(n \cdot \ell^2)$ and so is the complexity of the whole algorithm. The space complexity of the algorithm is determined by the size of the search tree, which is $\Theta(n \cdot \ell)$.

Recall that during the sorting step we remove all duplicates. Thus this algorithm is robust in the sense that it does not assume that all input vectors are distinct and the same complexity bound holds even if $X$ is a multiset. 

\section{GPU Implementation Issues} \label{sec:implissues}

In this section we discuss the implementation details of parallel algorithms described in the previous section. We shall omit an introduction to the computational model of GPGPU. The readers, who are not familiar with GPGPU programming, should refer to CUDA C literature \cite{cuda-best-practices,kirk2012programming}.
There are several limitations of GPU devices which are important from the algorithmic point of view. We are interested in algorithms which are able to:
(1) use coalesced memory access,
(2) maximize multiprocessor occupancy,
(3) hide memory latency.

\subsection{Heuristic algorithm}

In order to achieve high processor occupancy we need to define the number of blocks which is at least three or four times higher than the number of streaming processors. Memory latency may be hidden if there is sufficient number of warps assigned to the same processor and memory accessing is interspersed with computations. 

Algorithms \ref{alg:par-compute-dist} and \ref{alg:par-heuristic} contain two nested loops iterating over an array of results (it is an upper-triangular square array with zeros on the main diagonal). Using blocks as the parallel computation units in the outer loop and threads in the inner one gives us a fair number of blocks and threads achieving good occupancy and hiding memory latency. Each thread reads parts of two vectors into registers and then performs comparison. Thus a significant number of computational instructions are executed between reads and writes. 

Coalesced memory access is automatic if each vector is stored as a continuous array of bytes. 
Fragmented results of the comparison of two vectors in Algorithm \ref{alg:par-heuristic} (one array for each block) may be stored in a shared memory and added up in parallel by threads of this block using classical parallel reduction pattern.

\subsection{Tree-based algorithm}

Tree construction in Algorithm \ref{alg:par-construct} requires a synchronization after each level. Such a global synchronization can only be achieved by finishing a kernel and launching a new one. 
The number of threads in each kernel execution is equal to number of tree nodes in the previous level times $2^r$ (in our experiments we used $r=8$, so $2^r = 256$). All threads run independently and their division into blocks may be set arbitrarily in order to achieve best processor occupancy.

Algorithm \ref{alg:par-treebased} again contains two nested loops. The outer one is executed for each input vector and the inner one iterates over its coordinates. Similarly as in Algorithm \ref{alg:par-heuristic}, assigning the outer loop to blocks and the inner one to threads gives good parallelism properties. Each thread performs tree searching and reads in random memory locations. Coalesced memory reads are thus not possible. However, threads may still benefit from the global memory cache since up to $r$ threads may read the same byte from the memory performing independent searches.


\section{Experimental Results and Discussion}
\label{sec:experiments}

In order to evaluate our parallel algorithms we utilized the sample generator developed by Dobrowolski \cite{DBLP:conf/sii/Dobrowolski13} and some real-life scenes. The experiment was performed on a professional computation server (Intel Xeon E5-2620 2GHz, 15MB cache, 6 cores, 32GB RAM) equipped with NVIDIA Tesla K40 computational unit (2668 cores, 12GB memory). 

\begin{figure}[h]
\begin{center}
\includegraphics[width=.5\textwidth]{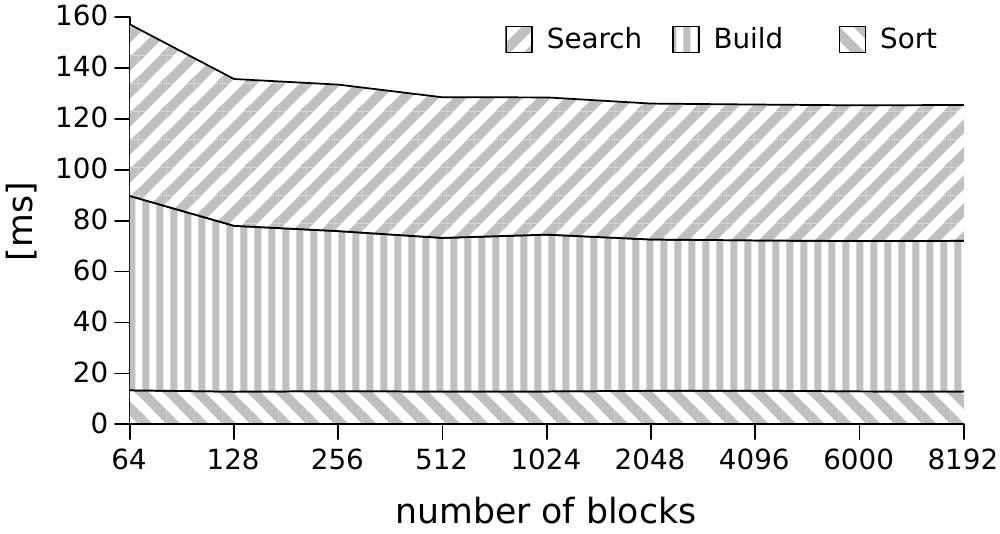}
\end{center}
\caption{\label{fig:blocks} The analysis of different block sizes for the tree-based algorithm. }
\end{figure}

In all parallel algorithms there are parameters which may influence their performance, i.e. the number of blocks and threads and the value of $h$ in the heuristic algorithm. According to NVIDIA white papers, due to the complication of the parallel processing model, the only way to find optimal values of these parameters for different devices and environments is to perform experiments. In the case of the value of $h$ (for the heuristic algorithm) our tests indicated that the optimal value for the CPU is 3, while for K40 it is 5. The rest of the experiments for heuristic algorithm were performed with these settings. An analysis of the size of the kernel grid for the parallel tree-based algorithm (divided into three stages: sorting, building and searching) is presented in Figure \ref{fig:blocks}. The total processing time was minimal for 4096 blocks of 32 threads. 

\begin{figure}[h]
\includegraphics[width=.5\textwidth]{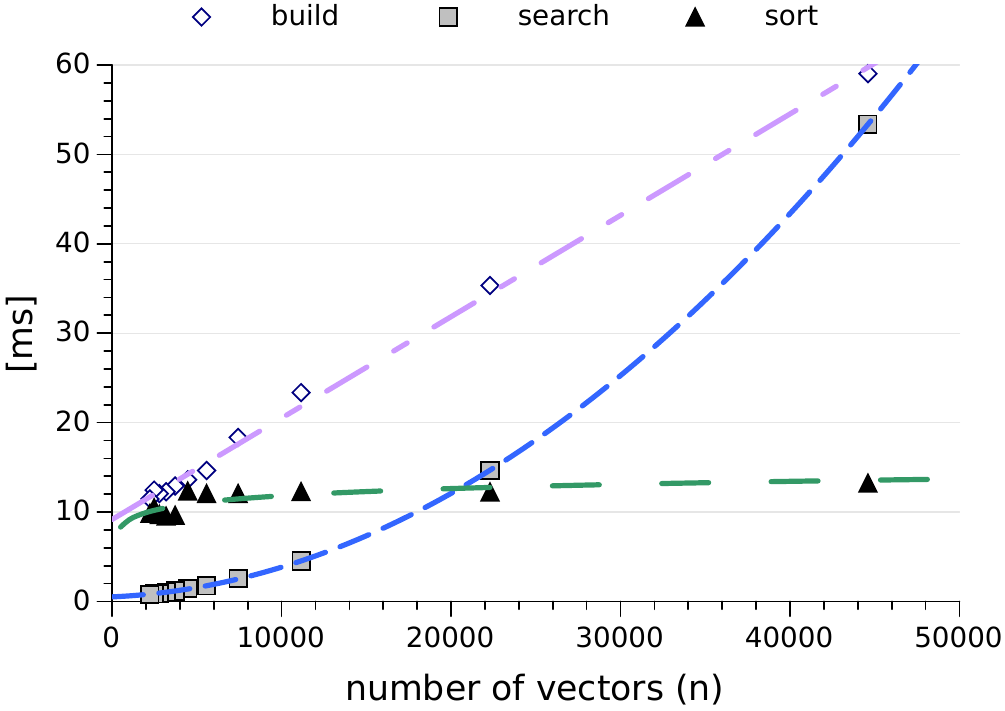}
\includegraphics[width=.5\textwidth]{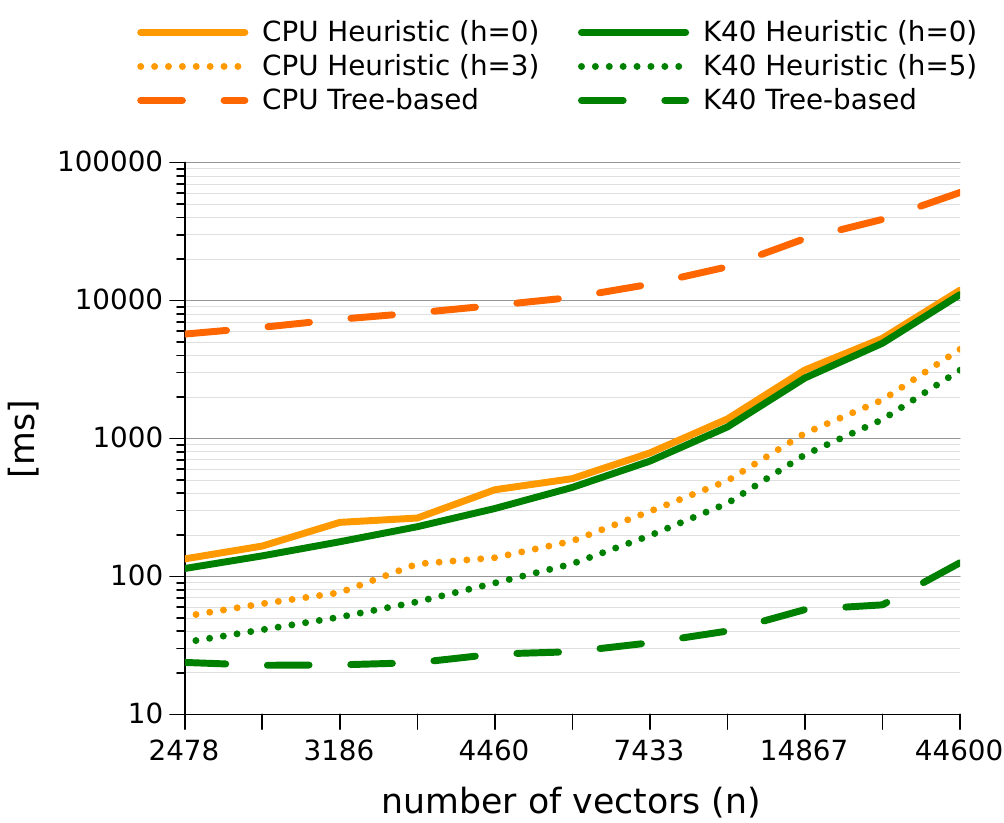}
\caption{\label{fig:results} The processing time of three stages of the tree-based algorithm (left). The processing time for sequential and parallel algorithms (right, note the logarithmic scale).}
\end{figure}

Figure \ref{fig:results} (left) presents the evaluation of the processing time in three stages of the tree-based algorithm: sorting, tree building, and neighbor searching. We can clearly see that searching time is growing faster than building, which is related to the $n^2$ factor in the complexity bound. However, for 44.000 vectors it is still smaller than the building part, due to high constants in the latter. 
Experiments show that the sorting stage does not influence the total processing time by more than 30\% in case of bigger input sets. 

On the right of the Figure \ref{fig:results} a comparison of several solutions is presented. 
Let us first analyze the heuristic methods. The sequential solution for the optimal value of $h$ (equal to 3) is significantly faster than the solution with $h$ set to 0, which corresponds to the naive solution. 
Similarly, parallel version with $h$ set to the optimal value (5) is much faster than the naive one. Both parallel and sequential procedures show similar growth of processing time for increasing size of the input data set. This shows that the algorithm scales well.
The best performance is achieved by the parallel tree-based procedure. Sequential version behaves similarly but significantly (more than two orders of magnitude) slower. 

\section{Conclusion and further research directions} \label{sec:future}

We presented two important parallel algorithms for construction of the cell graph in the motion planning problem. Our experiments show that the parallel solution based on a search tree is much faster than its sequential counterpart. This is also one of rare efficient search tree structures for GPU processors. We show that creating and searching such a structure can be efficient also on a SIMD-like processors, which were so far identified with vector processing. This was possible due to proper tree node construction and memory caching available in modern devices.

As a modification of the tree-based algorithm, Dobrowolski presented an algorithm, constructing the cell graph in $O(n \cdot \ell)$ time. Using an auxiliary data structure, the searching step can be performed in $O(n \cdot \ell)$ time.
Unfortunately, the experiments on the real data (see Section \ref{sec:experiments}) show that constructing the tree takes the majority of the execution time. Moreover, as this improved searching procedure requires lots of synchronization, it may actually lead to worse execution time.
A very natural research direction is to design a scalable parallel algorithm, constructing the cell graph for a given set of vectors in time $O(n \cdot \ell)$.

As mentioned before, cell graphs are used in motion planning. A path in the cell graph corresponding to a given scene is equivalent to some approximate solution of the motion planning problem.
There are several approaches to traversing large graphs using GPGPU (see \cite{conf/ppopp/MerrillGG12,kaczmarski2015improving}). An interesting problem is to design such an algorithm, taking into consideration its structure.


\end{document}